\documentclass[twocolumn]{aastex61}

\submitjournal{ApJ}

\shorttitle{Hydrodynamic simulations of propellers}
\shortauthors{Sei{\ss} et al.}
\begin{document}

\title{Hydrodynamic simulations of moonlet-induced propellers in Saturn's rings: Application to Bl\'eriot}

\correspondingauthor{Martin Sei{\ss}}
\email{martin.seiss@uni-potsdam.de}

\author{Martin Sei{\ss}}
\affiliation{Department of Physics and Astronomy, University of Potsdam, Potsdam, Germany}

\author{Nicole Albers}
\affiliation{LASP, University of Colorado, Boulder, USA} 

\author{Miodrag Srem{\v c}evi{\'c}}
\affiliation{LASP, University of Colorado, Boulder, USA} 

\author{J\"urgen Schmidt}
%\affiliation{Department of Physics and Astronomy, University of Potsdam, Potsdam, Germany}
\affiliation{Astronomy Research Unit, University of Oulu, Finland}

\author{Heikki Salo}
\affiliation{Astronomy Research Unit, University of Oulu, Finland}

\author{Michael Seiler}
\affiliation{Department of Physics and Astronomy, University of Potsdam, Potsdam, Germany}

\author{Holger Hoffmann}
\affiliation{Department of Physics and Astronomy, University of Potsdam, Potsdam, Germany}

\author{Frank Spahn}
\affiliation{Department of Physics and Astronomy, University of Potsdam, Potsdam, Germany}

%% Note that the \and command from previous versions of AASTeX is now
%% depreciated in this version as it is no longer necessary. AASTeX 
%% automatically takes care of all commas and "and"s between authors names.

%% AASTeX 6.1 has the new \collaboration and \nocollaboration commands to
%% provide the collaboration status of a group of authors. These commands 
%% can be used either before or after the list of corresponding authors. The
%% argument for \collaboration is the collaboration identifier. Authors are
%% encouraged to surround collaboration identifiers with ()s. The 
%% \nocollaboration command takes no argument and exists to indicate that
%% the nearby authors are not part of surrounding collaborations.

%% Mark off the abstract in the ``abstract'' environment. 

\begin{abstract}
One of the biggest successes of the Cassini mission is the detection of small moons (moonlets) embedded in Saturn’s rings which cause S-shaped density structures in their close vicinity, called propellers \citep{SpahnSremcevic2000,Tiscareno2006Natur,Sremcevic2007Natur}. Here, we present isothermal hydrodynamic simulations of moonlet-induced propellers in Saturn's A ring which denote a further development of the original model \citep{SpahnSremcevic2000}. We find excellent agreement between these new hydrodynamic and corresponding N-body simulations. Furthermore, the hydrodynamic simulations confirm the predicted scaling laws \citep{SpahnSremcevic2000} and the analytical solution for the density in the propeller gaps \citep{SremcevicSpahnDuschl2002}. Finally, this mean field approach allows us to simulate the pattern of the giant propeller Bl\'eriot, which is too large to be modeled by direct N-body simulations. Our results are compared to two stellar occultation observations by the Cassini Ultraviolet Imaging Spectrometer (UVIS), that intersect the propeller Bl\'eriot. Best fits to the UVIS optical depth profiles are achieved for a Hill radius of $590 \, \text{m}$, which implies a moonlet diameter of about $860 \, \text{m}$. Furthermore, the model favours a kinematic shear viscosity of the surrounding ring material of $\nu_0 = 340 \,\text{cm}^2 \, \text{s}^{-1}$, a dispersion velocity in the range of  $0.3  \, \text{cm} \, \text{s}^{-1} < c_0 < 1.5 \, \text{cm} \, \text{s}^{-1}$, and a fairly high bulk viscosity $7 < \xi_0/\nu_0 < 17$. These large transport values might be overestimated by our isothermal ring model and should be reviewed by an extended model including thermal fluctuations.
\end{abstract}

%% Keywords should appear after the \end{abstract} command. 
%% See the online documentation for the full list of available subject
%% keywords and the rules for their use.

%% From the front matter, we move on to the body of the paper.
%% Sections are demarcated by \section and \subsection, respectively.
%% Observe the use of the LaTeX \label
%% command after the \subsection to give a symbolic KEY to the
%% subsection for cross-referencing in a \ref command.
%% You can use LaTeX's \ref and \label commands to keep track of
%% cross-references to sections, equations, tables, and figures.
%% That way, if you change the order of any elements, LaTeX will
%% automatically renumber them.

%% We recommend that authors also use the natbib \citep
%% and \citet commands to identify citations.  The citations are
%% tied to the reference list via symbolic KEYs. The KEY corresponds
%% to the KEY in the \bibitem in the reference list below. 

%%%%%%%%%%%%%%%%%%%%%%%%%%%%%%%%%%%%%%%%%%%%%%%%%%%%%%%%%%%%%%%%%%%%%%%%%%
\section{Introduction}
%%%%%%%%%%%%%%%%%%%%%%%%%%%%%%%%%%%%%%%%%%%%%%%%%%%%%%%%%%%%%%%%%%%%%%%%%%

Saturn's dense rings consist of icy particles with sizes of centimeters up to tens of meters. Apart from this main population, small moons are embedded in these rings, which induce structures in the ring density by their gravitational influence. The largest examples are the two ring-moons Pan and Daphnis with radii of about 14 and 4 km, respectively \citep{Showalter1991,Porco2005IAUC}. Pan and Daphnis are massive enough to open and maintain a circumferential gap around their orbit \citep{Henon1981,Lissauer1981,PetitHenon1988,SpahnWiebicke1989,SpahnSponholz1989}. They further cause
wavy gap edges and corresponding wakes \citep{CuzziScargle1985,Showalter1986,Borderies1989Icar,Spahn1994Icar,Hertzsch1997AuA,LewisStewart2000,Weiss2009AJ,Seiss2010Icarus}. Pan additionally maintains a central ringlet on its orbit \citep{Dermott1980Natur,Dermott1981Icar,SpahnSponholz1989}.

Smaller moons (moonlets) with radii between 50 m and 500 m cannot be observed directly, but they reveal themselves by much larger S-shaped density structures in their vicinity, called propellers, caused by their gravitational interaction with the surrounding ring material \citep{SpahnSremcevic2000,SremcevicSpahnDuschl2002}. To date more than 150 propellers have been detected in the A ring \citep{Tiscareno2006Natur,Sremcevic2007Natur,Tiscareno2008AJ}. The propeller named Bl\'eriot is the largest example, large enough to be tracked over a longer time span in Cassini images. Analysis of the moonlets' orbital motion revealed an unexplained wandering along its longitude relative to the motion expected from the Keplerian angular speed
% induced by the gravity of the planet 
\citep{Tiscareno2010ApJ,Seiler2017ApJ}.

First theoretical investigations describing the density structure in the vicinity of a small, embedded moonlet in the rings were made by \citet{SpahnSremcevic2000}. They used a model that combines gravitational scattering of ring particles in the vicinity of the moon tending to open a gap in the rings, and diffusion tending to a closing of the gap. This model predicts a structure consisting of two radially shifted gaps with limited azimuthal extent. Furthermore, \citet{SpahnSremcevic2000} derived scaling laws which indicate that (a) the radial dimension of the gaps is directly proportional to the Hill radius $h$ of the moon
\begin{equation}
	h = a \left( \frac{M_{\rm m}}{3 \, M_{\rm p}} \right)^{1/3}
\end{equation}
and (b) the azimuthal extent scales with $h^3/\nu_0 \propto M_{\rm m}/\nu_0$. Here, the semi-major axis of the moon is denoted by $a$, and its mass by $M_m$, whereas the mass of the planet is labeled by $M_{\rm p}$. \citet{SremcevicSpahnDuschl2002} presented an analytical solution for the gap density profile. An extension of the model to the vertical degree of freedom was developed by \citet{Hoffmann2013ApJ,Hoffmann2015Icar}, which explained the shape of the shadow cast by the propeller Earhart on the rings in Cassini ISS images taken close to Saturn's vernal equinox in August 2009.

\citet{Seiss2005GeoRL} were the first to employ N-body simulations in order to investigate a moonlet-induced propeller structure. This approach allowed for simultaneous modeling of the gap and the associated wake structures, and the results confirmed the predicted scaling laws. Further, N-body simulations which included self-gravity \citep{Sremcevic2007Natur,LewisStewart2009Icar} as well as a ring particle size distribution \citep{LewisStewart2009Icar} were published after the first propellers were discovered. Especially for small propeller moons, self-gravity wakes are expected to interact with the wake pattern induced by the moonlets, potentially destroying the moonlet wakes completely. Further, the migration of the moonlet orbit was studied by \cite{LewisStewart2009Icar} and \citet{Rein2010}.

While N-body simulations are a powerful tool to investigate the formation of propellers, they are computationally expensive, so that usually only the close vicinity of a small moonlet can be simulated. The problem is that the size of the simulated ring particles has to be close to the effective particle size in the rings (of the order of 1 meter) to get the right macroscopic properties of the ring as for example velocity dispersion, viscosity and pressure. Thus, with growing simulation area the number of simulated particles becomes prohibitively large\footnote{In the self-gravitating N-body models in \citet{Sremcevic2007Natur}, 345,000 particles were needed to simulate a moonlet with a radius of 20 m.}. In order to lift this limitation, we use the hydrodynamic approach with values for pressure and transport coefficients of the granular ring matter determined from N-body simulations \citep{SaloSchmidtSpahn2001,Daisaka2001} or directly from observations \citep{Tiscareno2007Icar,Sremcevic2008DPS}. This is the first time hydrodynamic simulations are used to simulate propellers, whereas this numerical method is an established tool to investigate the evolution of planetary embryos in a pre-planetary disk \citep[e.g.][]{Kley1999MNRAS,Lubow1999ApJ}, a system physically similar to moonlets embedded in a planetary ring.

In the following, the hydrodynamic equations are introduced in Section \ref{sec:HydroEq} and the simulation method is sketched in Section \ref{sec:methods}. Results are presented and discussed in Sections \ref{sec:results} and \ref{sec:conclusion}.

%%%%%%%%%%%%%%%%%%%%%%%%%%%%%%%%%%%%%%%%%%%%%%%%%%%%%%%%%%%%%%%%%%%%%%%%%%
\section{Hydrodynamic equations}
\label{sec:HydroEq}
%%%%%%%%%%%%%%%%%%%%%%%%%%%%%%%%%%%%%%%%%%%%%%%%%%%%%%%%%%%%%%%%%%%%%%%%%%

Physically, Saturn's main rings are formed by icy particles %form a dense granular gas 
orbiting around the planet. In our simulations the flow of these ring particles is treated as a granular gas, perturbed by a small moonlet embedded in the disk. The main rings have a very small vertical extension of 10 to 100 m, especially in comparison to the diameter of about 270,000 km. Thus, it is convenient for many purposes to describe the rings with a 2 dimensional model using vertically integrated quantities (see \citet{Spahn2018book} and references therein).

The evolution of surface mass density $\Sigma$ and flux $\Sigma {\bf v}$ of the ring material in a co-rotating frame are described by the continuity equation
\begin{equation}
\partial_t \Sigma + {\bf \nabla} \cdot \Sigma {\bf v}  = 0
\end{equation}
and the momentum balance %Navier-Stokes equation
\begin{equation}
\partial_t \Sigma {\bf v} + \nabla \cdot (\Sigma {\bf v} \circ {\bf v}) 
       = -\Sigma \, \nabla \cdot  (\Phi_{\rm p} + \Phi_{\rm m}) + {\bf f_{\rm i}}
       - {\bf {\nabla}} \cdot {\sf \hat{P}} \\
\end{equation}
written here in the flux conserved form. The symbol $\circ$ denotes the dyadic product and vectors are marked by bold letters. Further, the gravitational potentials of central planet and moonlet are labeled by $\Phi_{\rm p}$ and $\Phi_{\rm m}$, respectively. The inertial forces in the co-rotating frame are the centrifugal and Coriolis force
\begin{equation}
{\bf f_{\rm i}} = - \Sigma \, {\bf \Omega} \times ({\bf \Omega} \times {\bf r}) 
	- 2 \Sigma \, {\bf \Omega} \times {\bf v} \,
\end{equation}
where ${\bf \Omega}$ denotes the Kepler frequency of the moonlet. We neglect higher gravitational moments of Saturn, treating it as a spherical planet. More realistic would be the assumption of an oblate body, but the main effect would be a change in the Keplerian, epicylic and vertical frequencies by less than one per cent - a small effect which is neglected for simplicity.

The pressure tensor ${\sf \hat{P}}$ can be described with the Newtonian ansatz as
\begin{equation}
P_{ij} = p \, \delta_{ij} - \Sigma \nu  \left( \frac{\partial v_i}{\partial x_j}
  + \frac{\partial v_j}{\partial x_i} \right)
  + \Sigma \left( \frac{2}{3} \, \nu - \xi \right) \, {\bf {\nabla}} \cdot {\bf v} \,\delta_{ij}
\label{Eq:PressureTensor}
\end{equation}
where $p$, $\nu$ and $\xi$ denote the scalar pressure as well as the kinematic shear and bulk viscosities, respectively. The pressure and the viscosities depend on the local density, and are parameterized by power laws \citep{Spahn2000Icar,SaloSchmidtSpahn2001}
\begin{eqnarray}
\label{Eq:p}
p & = & p_0 \, \left( \frac{\Sigma}{\Sigma_0} \right)^{\alpha} \\
\nu & = & \nu_0 \, \left( \frac{\Sigma}{\Sigma_0} \right)^{\beta} \, .
\label{Eq:nu}
\end{eqnarray}
Furthermore, the ratio between shear and bulk viscosity is set constant
\begin{equation}
\xi = \frac{\xi_0}{\nu_0} \, \nu \, .
\end{equation}
In the simplest case, the unperturbed pressure is given by the ideal gas relation $p_0=\Sigma_0 \, c_0^2$, where $c_0$ denotes the dispersion velocity. We use an isothermal model; thus, the dispersion velocity $c_0$ and the related granular temperature $T=c_0^2/3$ are constant.

The Newtonian relation (Eq. \ref{Eq:PressureTensor}) constitutes a linear response of the pressure tensor on the change in the shear. This assumption might be violated in regions with larger changes in the velocity field or in the surface mass density. Furthermore, our parameterization of the viscous effects may be an over-simplification, which does not include all aspects of the kinetic processes happening in the rings. For example, dissipative processes within propellers may not take the form of a single 'viscosity' that can be
directly applied to describe other ring phenomena such as the long-term evolution of Saturn's dense rings \citep{Charnoz2010Natur},
%self-gravity wakes \citep{Daisaka2001}, 
spiral density waves \citep{Tiscareno2007Icar}, or the clearing of a gap \citep{Tajeddine2017ApJS,Tajeddine2017Icar}. Nevertheless, interpreting with care our viscosity parameter remains a useful way to describe dissipative processes within our propeller simulations and is related to the transport of angular momentum in unperturbed rings, as we show with our comparison to N-body simulations in Section \ref{CompNbodyHydro}.

%%%%%%%%%%%%%%%%%%%%%%%%%%%%%%%%%%%%%%%%%%%%%%%%%%%%%%%%%%%%%%%%%%%%%%%%%%
\section{Method}
\label{sec:methods}
%%%%%%%%%%%%%%%%%%%%%%%%%%%%%%%%%%%%%%%%%%%%%%%%%%%%%%%%%%%%%%%%%%%%%%%%%%

Propellers are small objects in comparison to the ring dimensions, and thus, just a small region of the ring is simulated with the origin of the reference frame fixed at the position of the moonlet on a circular orbit. The acting forces can be linearized, leading to the Hill problem \citep{Hill1878}. In the following, $x$ and $y$ denote radial and azimuthal distance from the moonlet, respectively. Positive $y$ is in the direction of orbital motion and positive $x$ points away from the central planet. The simulation program calculates the perturbed flux $\Sigma {\bf u}$ using the velocity ${\bf u} = {\bf v} + (3/2) \Omega \, x \, {\bf e_y}$, where ${\bf u}$ is the mean velocity ${\bf v}$ reduced by the Keplerian shear velocity. The systematic shear velocity $-(3/2) \Omega \, x$ arises from the linearized radial dependence of the Kepler velocity. The Kepler frequency is calculated by $\Omega=\sqrt{GM_{\rm p}/a^3}$ at the radial location of the moonlet $a$, where the gravitational constant is labeled by $G$. Introducing $\Sigma {\bf u}$ simplifies the equations and increases the stability of the advection scheme described below. Thus, the Navier-Stokes equation now reads
\begin{equation}
\partial_t \Sigma {\bf u} + \nabla \cdot (\Sigma {\bf u} \circ {\bf v}) 
       = -\Sigma \, \nabla \cdot  \Phi_{\rm m} + {\bf f_{\rm T}}
       - {\bf {\nabla}} \cdot {\sf \hat{P}}  \, \, \, .
\end{equation}
The remaining inertial forces in this system can be written in the form 
\begin{equation}
	{\bf f_{\rm T}}= 2 \, \Omega \, \Sigma u_y \,{\bf e_x}
		-\frac{1}{2} \, \Omega \, \Sigma u_x  \,{\bf e_y} \, .
\end{equation}

The gravitational potential of the moon
\begin{equation}
	\Phi_m = - \frac{G M_{\rm m}}{\sqrt{x^2+y^2+\epsilon^2}}
\end{equation}
is modified by a smoothing radius $\epsilon$, softening the gravitational potential in the close vicinity of the moon center. Usually a value $\epsilon=0.2 h$ is chosen - small enough not to influence the final results.

The calculation region is a rectangular cutout of the ring with dimensions $(x_{\rm min},x_{\rm max})$ and $(y_{\rm min},y_{\rm max})$ centered at the position of the moonlet. For the numerical intergration the region is discretized into $N_x \times N_y$ equal-sized cells. The complete set of equations is integrated until a steady state is established. We applied the method of directional operator splitting to solve separately the hydrodynamic equations for the $x$ and $y$ direction \citep{Strang1968SJNA}. This has the big advantage that the code can easily be parallelized. The advection term is solved with the first order donor-cell algorithm \citep{LeVeque2002}, which is easy to implement, but induces an artificial diffusion. For this reason, we make sure that the resolution is sufficiently high and the results are not affected significantly. For the simulations of Bl\'eriot we use a second order scheme with MinMod flux limiter \citep{LeVeque2002} in order to better conserve the wake crests. The flux term ${\bf {\nabla}} \cdot {\sf \hat{P}}$, representing the effects of pressure and viscous transport, is integrated forward in time with an explicit scheme. 

The advective time step is chosen as
\begin{equation}
	\Delta t_{\rm ad} = \frac{1}{2} \, 
		\min \left( \frac{\Delta x}{(|v_x|+c_0)} , \frac{\Delta y}{(|v_y|+c_0)} \right)
\end{equation}
to fulfill the Courant-Friedrichs-Lewy condition, necessary for a stable algorithm, where $\Delta x$ and $\Delta y$ denote the radial and azimuthal dimension of a simulation cell, respectively. The viscous time step is calculated from
\begin{equation}
	\Delta t_{\rm vis} = \min \left( \frac{\Delta x^2}{\nu} , \frac{\Delta y^2}{\nu} \right) \, .
\end{equation}
Usually the advective time step is much smaller than the viscous time step, and thus, sets the integration time step in our simulations.

Boundary conditions of the calculation region are chosen such that perturbations flow freely out of the box, while the inflow is unperturbed. This is especially important at the azimuthal boundaries, where due to Kepler shear, material is flowing into the box at $x<0,y=y_{\rm min}$ and $x>0,y=y_{\rm max}$ and flowing out at $x<0,y=y_{\rm max}$ and $x>0,y=y_{\rm min}$. The moonlet perturbs the ring by its gravity. The softening induced with the parameter $\epsilon$ allows us to let ring matter flow through the space occupied physically by the moonlet without any significant effect on the shape of the forming propeller. We also tested the influence of reflecting boundary conditions on the moons surface at radius $R_m$, but did not find a significant influence on the propeller pattern.

In the simulation code time is scaled by $1/\Omega$ and length by the Hill radius $h$. This means that also pressure and viscosities are scaled accordingly, although they do not depend on the moonlet mass (Hill radius). These may be converted to SI units in applications by choosing a value of the Hill radius. Self-gravity of the ring material is neglected in our simulations; thus, we scale the density by a constant $\Sigma_0$ because only the relative change of the density is needed to calculate pressure and viscosities (see Equations \ref{Eq:p} and \ref{Eq:nu}).

%%%%%%%%%%%%%%%%%%%%%%%%%%%%%%%%%%%%%%%%%%%%%%%%%%%%%%%%%%%%%%%%%%%%%%%%%%
\section{Results}
\label{sec:results}
%%%%%%%%%%%%%%%%%%%%%%%%%%%%%%%%%%%%%%%%%%%%%%%%%%%%%%%%%%%%%%%%%%%%%%%%%%

\begin{figure*}[ht!]
\plotone{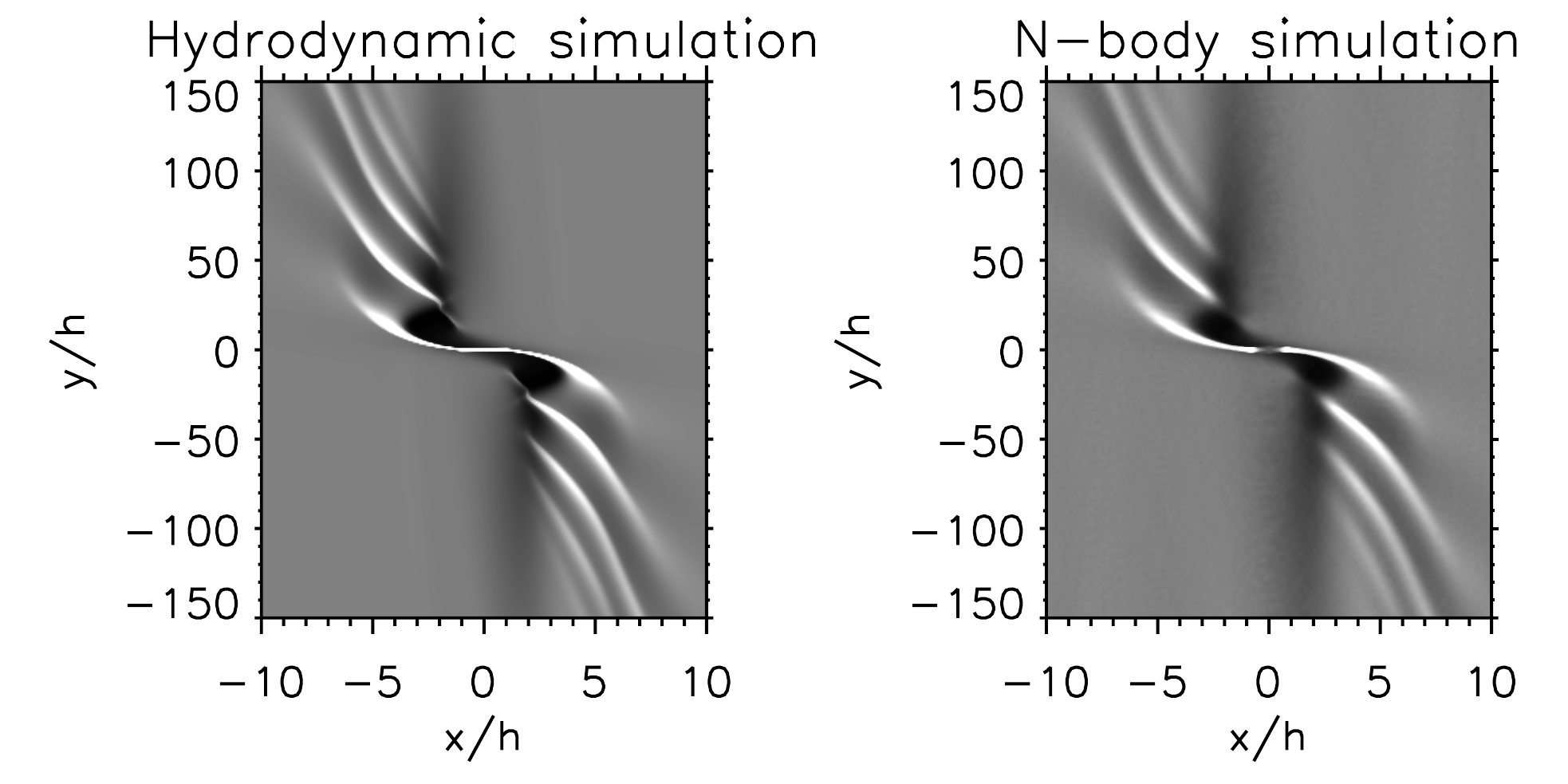}
  \caption{Propeller structure around a moonlet located in the center of the box. Axes are scaled by the Hill radius $h$, where $x$ and $y$ denote the radial and the azimuthal directions, respectively. The left panel presents the surface density from a hydrodynamic simulation, whereas the right panel shows the resulting surface density from a N-body simulation of a moonlet with $h = 19.6 \, \text{m}$ composed from 50 averaged snapshots taken every 0.1 orbits, after 20 orbital periods when a steady state has been reached. Parameters for the hydrodynamic simulation are chosen according to the N-body simulation.}
  \label{fig:HydroNbody}
\end{figure*}

The surface mass density distribution resulting from a hydrodynamic propeller simulation is plotted in the left panel of Figure \ref{fig:HydroNbody}, where the moonlet in the box center has a Hill radius of 19.6 m. The simulation region used here is divided into $N_x \times N_y = 1200 \times 4000$ cells and it extends from -15$h$ to 15$h$ in radial and -200$h$ to +200$h$ in azimuthal direction. A steady state is established after about 20 orbits. Parameters for the hydrodynamic simulations are set to $c_0=0.54 \, \text{mm} \, \text{s}^{-1}$, $\alpha=1.79$, $\nu_0=4.2 \, \text{cm}^2 \, \text{s}^{-1}$, $\xi_0=3 \nu_0$ and $\beta=0.67$. These are chosen according to results gained via N-body simulations by \citet{SaloSchmidtSpahn2001} for a ring of 1 meter sized particles with optical depth of $\tau=0.5$, with the \citet{Bridges1984Natur} elasticity law and an enhanced vertical frequency $\Omega_z/\Omega_0=3.6$, mimicking the effect of an enhancement of the collision frequency induced by self-gravity \citep{WisdomTremaine1988}. The viscosities and the pressure include the local and nonloal components, which arise from particle random motions and from mutual impacts, respectively \citep{WisdomTremaine1988}.

The density pattern shows the typical propeller fingerprint, with two radially displaced gaps that fade downstream having their density minima at about $x= \pm 2h$. Particles that pass the moon in larger radial distance are deflected just slightly towards the moon. This induces a coherent motion of the ring material and leads to the formation of wakes \citep{Showalter1986}.

%%%%%%%%%%%%%%%%%%%%%%%%%%%%%%%%%%%%%%%%%%%%%%%%%%%%%%%%%%%%%%%%%%%%%%%%%%
\subsection{Comparison with N-body simulations}
%%%%%%%%%%%%%%%%%%%%%%%%%%%%%%%%%%%%%%%%%%%%%%%%%%%%%%%%%%%%%%%%%%%%%%%%%%

\label{CompNbodyHydro}

In this section, we compare the density structure from the hydrodynamic simulation with that obtained from N-body simulations using the local code developed by \citet{Salo1995}. This code was applied to propellers by \citet{Seiss2005GeoRL} and it allows us to validate our hydrodynamic approach. This is not only important to judge the performance of our numerical scheme. It also allows us to test the validity of the hydrodynamic approximation for this problem, specifically the isothermal model used here and the parameterization of viscosities and pressure. 

%The hydrodynamic approach requires that both the mean free path and the size of the particles are much smaller than (i) the typical length scales of the macroscopic structure investigated and (ii) the epicyclic length of particle orbits. Therefore, the inverse collision frequency and the impact duration of particles should be much shorter than one orbital period. For rings, these assumptions are not always fulfilled, especially in regions of very low or very high densities \citep[see e.g. discussion in][]{Seiss2011MMNP}. Propellers are mainly located in the A ring, where particles have sizes between centimeters and tens of meters \citep[][and references therein]{Cuzzi2009book} and the mean free path is expected to be of the same order, leading to a mean collision frequency of about 10 collisions per orbit per particle\footnote{\citet{Hoffmann2015Icar} determined 6 to 11 collisions per orbit per particle on base of a model of the Earhart shadow, while N-body simulations with full self-gravity yield an order of magnitude larger frequencies \citep{Salo1995}.}. Since the propeller gaps are a few Hill radii wide in radial direction, it takes many orbital periods to refill the gaps with ring material. Thus, hydrodynamics should work well here, because mean free path and collision time are much smaller. On the other hand, moonlet wakes are winding up downstream the moon, getting very tight, along the azimuthal direction, and furthermore, particles need only one orbit to pass a single wake crest, so that a local thermodynamic steady-state may not always establish.

A hydrodynamic approach to ring dynamics is best justified if the mean free path of the ring particles is appreciably less than the epicyclic length. This condition is not always fulfilled in Saturn's rings, especially in regions of very low densities  \citep[see e.g. discussion in][]{Seiss2011MMNP}. Propellers are mainly located in the A ring. Here particles have sizes between centimeters and tens of meters \citep[][and references therein]{Cuzzi2009book} and optical depths are as high as 0.5 to 1. Therefore, a hydrodynamic model should work well to describe the propeller dynamics. For example, the propeller gaps are a few Hill radii wide in radial direction, which is way larger than the mean free path. On the other hand, moonlet wakes are winding up in azimuthal direction, downstream from the moon. Particles need only one orbit to pass a single wake crest, so that a local thermodynamic steady-state may not always establish.

For comparison with the hydrodynamical model we performed a N-body simulation with $N=598,000$ particles of radii $R=1 \, \text{m}$ in a box with size $L_x \times L_y = 476 \,\text{m} \, \times \, 7900 \, \text{m}$. The moonlet is located in the center of the simulation area having a radius of $R_m=15 \, \text{m}$. Together with the bulk density of ice $\varrho=910 \text{kg/m}^3$, the Hill radius of the moonlet becomes $h=19.6 \, \text{m}$ ($a= 100,000 \, \text{km}$). The geometric optical depth in the box, defined by
\begin{equation}
	\tau = \frac{\pi R^2 \, N}{L_x \, L_y} \, \, ,
\end{equation}
is then $\tau=0.5$ - a value representative for the later application to Bl\'eriot. The inelastic collisions are modeled with a velocity dependent normal coefficient of restitution introduced by \citet{Bridges1984Natur}
\begin{equation}
	\epsilon(g_n) = \left( \frac{g_n}{v_c} \right)^{-0.234}
\end{equation} 
accounting for the dissipation of energy during the collisions. The normal component of the relative velocity between the impact partners is denoted by $g_n$ and the scale parameter $v_c$ equals $0.077 \, \text{mm} \, \text{s}^{-1}$ \citep{Bridges1984Natur}.

\begin{figure*}[ht!]
\plotone{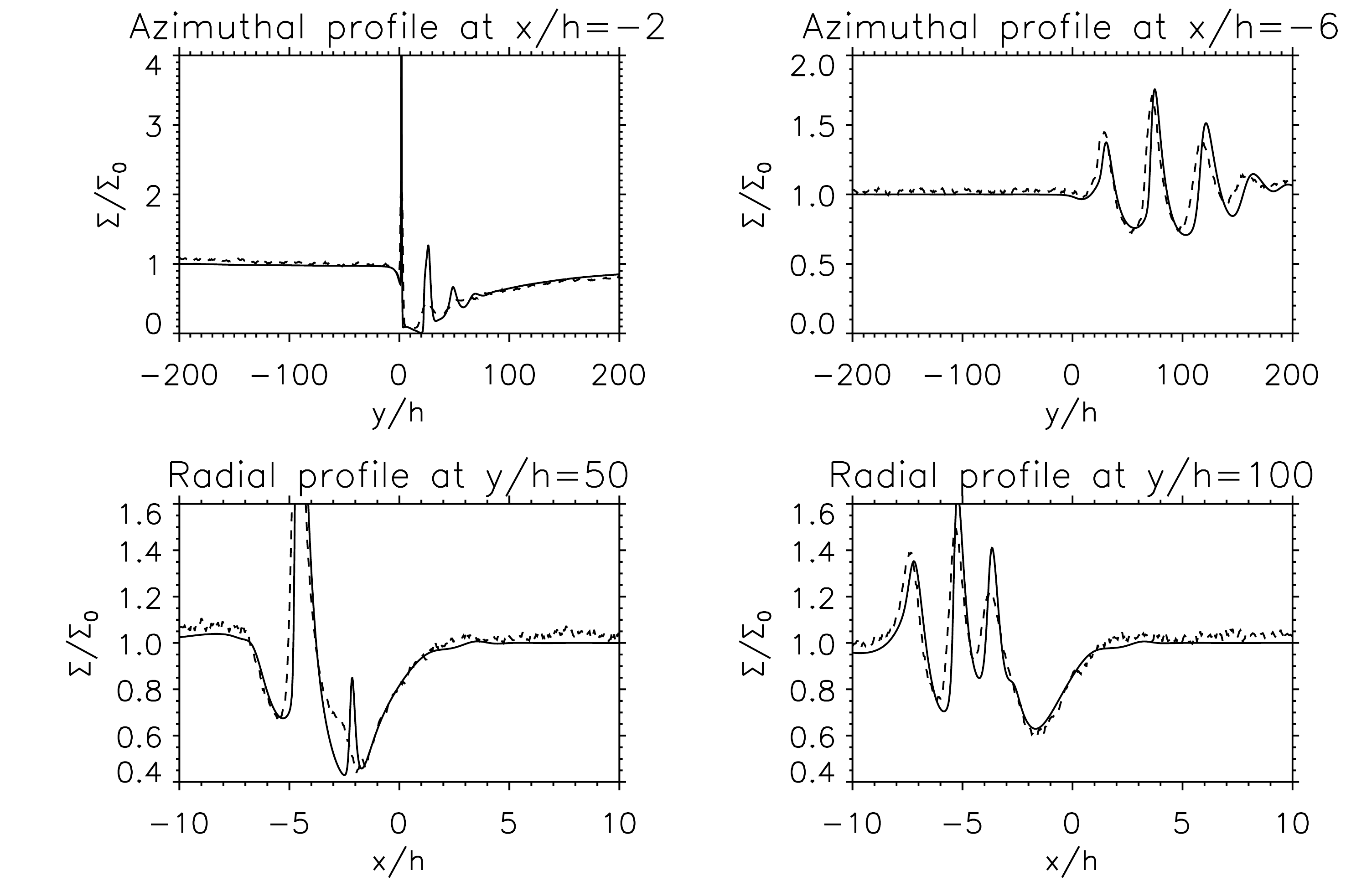}
  \caption{Density profiles of the hydrodynamic (solid lines) and N-body (dashed lines) simulations of a propeller shown in Figure \ref{fig:HydroNbody}. The profiles represent cuts along the azimuthal (upper panels) and the radial (lower panels) direction at various locations. The moonlet is located at $x=0$ and $y=0$.}
  \label{fig:HydroNbodyCuts}
\end{figure*}

The density pattern from the N-body simulation is presented in the right panel of Figure \ref{fig:HydroNbody}. For comparison, the left panel shows the hydrodynamic result, where pressure and transport coefficients are taken from N-body simulations \citep{SaloSchmidtSpahn2001}. Both results look nearly identical. A more detailed comparison of the profiles in Figure \ref{fig:HydroNbodyCuts} reveals an excellent agreement for the gap closing (upper left panel). Apart from boundary effects, profiles of the wakes also agree well (upper right panel) as long as the wake crests are not too closely spaced. Furthermore, the finite size of the ring particles smoothes the wake crests in the N-body simulation. However, this effect should play a smaller role when we apply the hydrodynamic model to the large propeller Bl\'eriot.

Summarizing, N-body and hydrodynamic simulations of propellers show a quite good agreement provided optical depth and moonlet size are sufficiently large. The good agreement of the gap profiles for N-body and hydrodynamic simulations shows that our concept of a viscosity parameter is reasonable for the hydrodynamic simulations if we apply the viscosity measured in the unperturbed N-body simulations to the hydrodynamic simulations. We therefore confidently apply the hydrodynamic code to large A-ring propellers orbiting between the Encke and Keeler gaps which are otherwise too large to be studied by N-body simulations. Note, self-gravity of the disk is omitted in our simulations. However, since self-gravity wakes are much smaller in size than the pattern of the Trans-Encke propellers, self-gravity will mainly affect the pressure and transport coefficients. 

%%%%%%%%%%%%%%%%%%%%%%%%%%%%%%%%%%%%%%%%%%%%%%%%%%%%%%%%%%%%%%%%%%%%%%%%%%
\subsection{Radial and azimuthal scaling of the propeller gap}
%%%%%%%%%%%%%%%%%%%%%%%%%%%%%%%%%%%%%%%%%%%%%%%%%%%%%%%%%%%%%%%%%%%%%%%%%%

We have seen that the outcome of N-body simulations and the hydrodynamic approach agree well. Next, we compare our results to the predictions by \citet{SpahnSremcevic2000} and \citet{SremcevicSpahnDuschl2002} in order to confirm further the applicability of our approach before we apply the hydrodynamic integrations to Bl\'eriot.

\begin{figure*}[ht!]
  \plotone{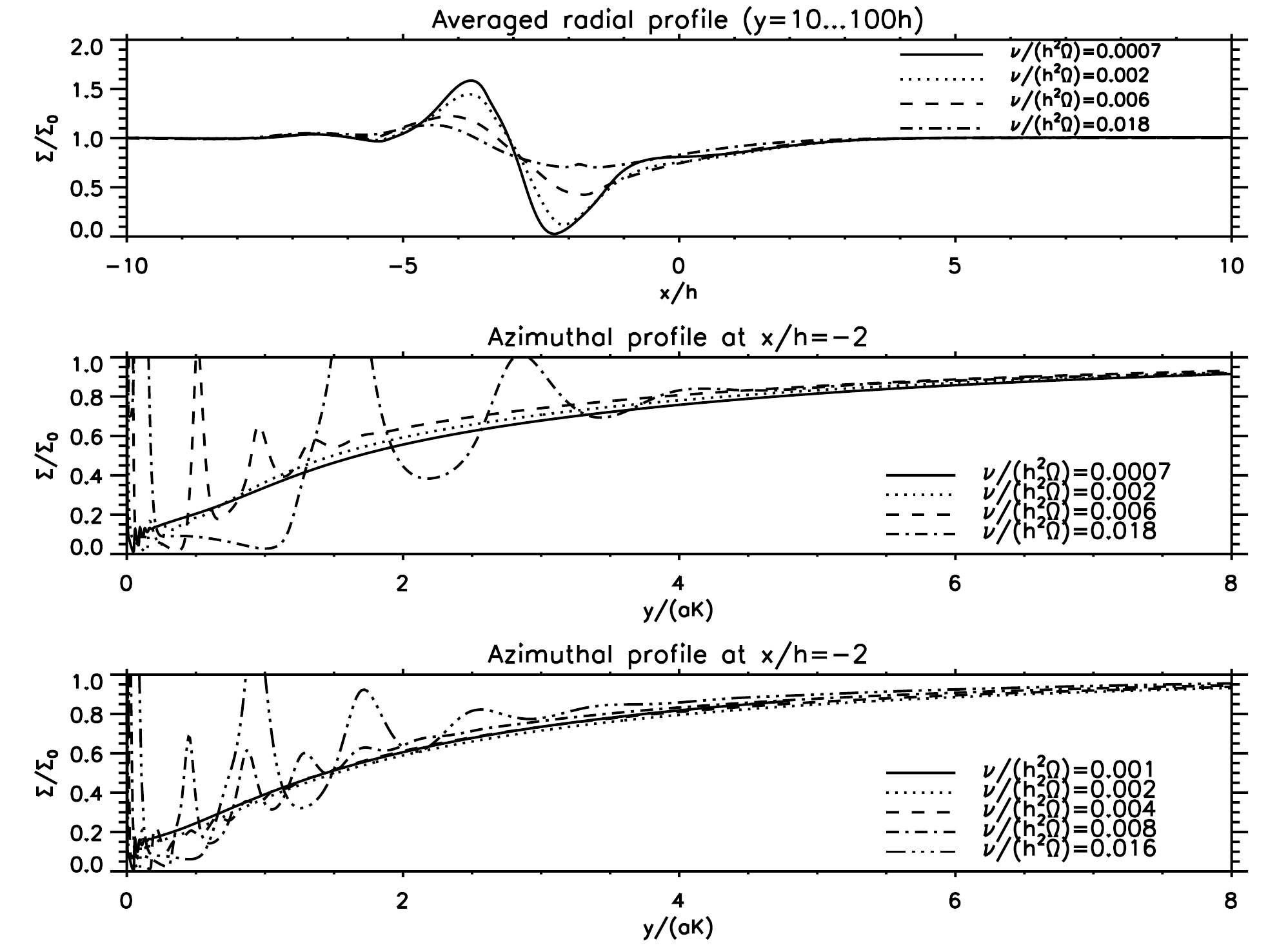}
  \caption{Comparison of gap profiles for hydrodynamic simulations with different shear viscosities. Upper and middle panel: radial and azimuthal cuts for fixed $\xi_0=3\nu_0$, $\beta=0.67$, $c_0=0.14 h \Omega$, $\alpha=2.15$. Lower panel: azimuthal cut for fixed $\xi_0=3\nu_0$, $\beta=0$, $c_0=0$.}
  \label{fig:HydroScalings}
\end{figure*}

The radial structure of the gap in the close vicinity of the moon is mainly caused by the gravitational scattering of the moonlet. The equations of motion can be scaled by the Hill radius $h$ if one neglects viscous transport and pressure. Thus, the Hill radius is the typical radial scale of the resulting density structure. This is demonstrated in Figure \ref{fig:HydroScalings} (top panel), where the azimuthally averaged radial profiles ($y/h$ from $0$ to $100$) from the hydrodynamic simulations are plotted using different values of the viscosity. In all simulations, gap minimum and adjacent maximum are located at about $x=-2h$ and $x=-4h$ for $y>0$, respectively. Different viscosity values change only the density level, but not significantly the location of minima and maxima. Insofar, the radial location of the gap minimum carries information about the mass of the moonlet.

\citet{SremcevicSpahnDuschl2002} showed that the azimuthal structure should scale with the diffusion length as
\begin{equation}
	a K = \frac{\Omega\, h^3}{2(1+\beta) \, \nu_0}   \, .
\end{equation}
where $K$ is a measure for the length of the induced propeller gap. The applicability of this expression is demonstrated in the middle and bottom panels of Figure \ref{fig:HydroScalings}, where the azimuthal profiles from simulations with different viscosities are plotted in scaled azimuthal coordinates. In the middle panel density dependent viscosities and pressure are used ($\beta=0.67$, $c_0=0.14 h \Omega$, $\alpha=2.15$), whereas in the bottom panel the simulations are performed with constant viscosity and zero pressure. The curves obey the azimuthal scaling very well in the region where the wakes are damped out, in particular having in mind that the viscosities differ by a factor of up to 81. It turns out that the density dependence of the viscosity in form of the power law parameter $\beta$ has only a small effect on the final profile, especially if one accounts for the $(1+\beta)$ factor in the scaling length $a K$, in which case the profiles nearly fall on top of each other. Thus, if the Hill radius has been derived from the radial profile, and if the wakes do not compromise the quality of the azimuthal profile fit, the azimuthal profile can be used to determine the parameter combination $(1+\beta) \, \nu_0$. 

%%%%%%%%%%%%%%%%%%%%%%%%%%%%%%%%%%%%%%%%%%%%
\subsection{Comparison to analytical model}
%%%%%%%%%%%%%%%%%%%%%%%%%%%%%%%%%%%%%%%%%%%%

The relaxation of the gap downstream from the moon can be described in terms of a linearized version of the diffusion equation \citep{SremcevicSpahnDuschl2002}
\begin{equation}
	\frac{\Omega_0}{2 \,(1+\beta)\, \nu_0} \partial_y \Sigma
 		=  - \frac{1}{x} \partial^2_x \Sigma  \, \, .
\label{DiffEq}
\end{equation} 
An approximate Green solution solving the problem for $\Sigma(x,y=0) = \delta(x - x_0)$ has been found \citep{SremcevicSpahnDuschl2002} in the form
\begin{eqnarray}
  G(x,y,x_0) & = & - \frac{\sqrt{3} x_0}{2h} \left( \frac{3 y}{a K} \right)^{-2/3}
     \exp \left[ \frac{a K \, (x^3+x_0^3)}{9 \, y \, h^3} \right] \nonumber \\
		& & \qquad \cdot \,  \mbox{Bi} \left[ \left( \frac{3 y}{a K} \right)^{-2/3} \frac{x_0 \, x}{h^2}  \right]
  \label{SGF}
\end{eqnarray}
where $x$ and $y$ scale with $h$ and $a K$, respectively, as discussed above. $\text{Bi}(z)$ denotes the Airy function. The general solution reads
\begin{equation}
	\Sigma(x,y) = \int \Sigma(x_0,y=0) \, G(x,y,x_0) \, dx_0
	\label{SGeneralSolution}
\end{equation}
and can then be computed from the radial profile $\Sigma(x_0,y=0)$ that can be found from the gravitational scattering by the moon. 

\begin{figure*}[htb!]
  \plotone{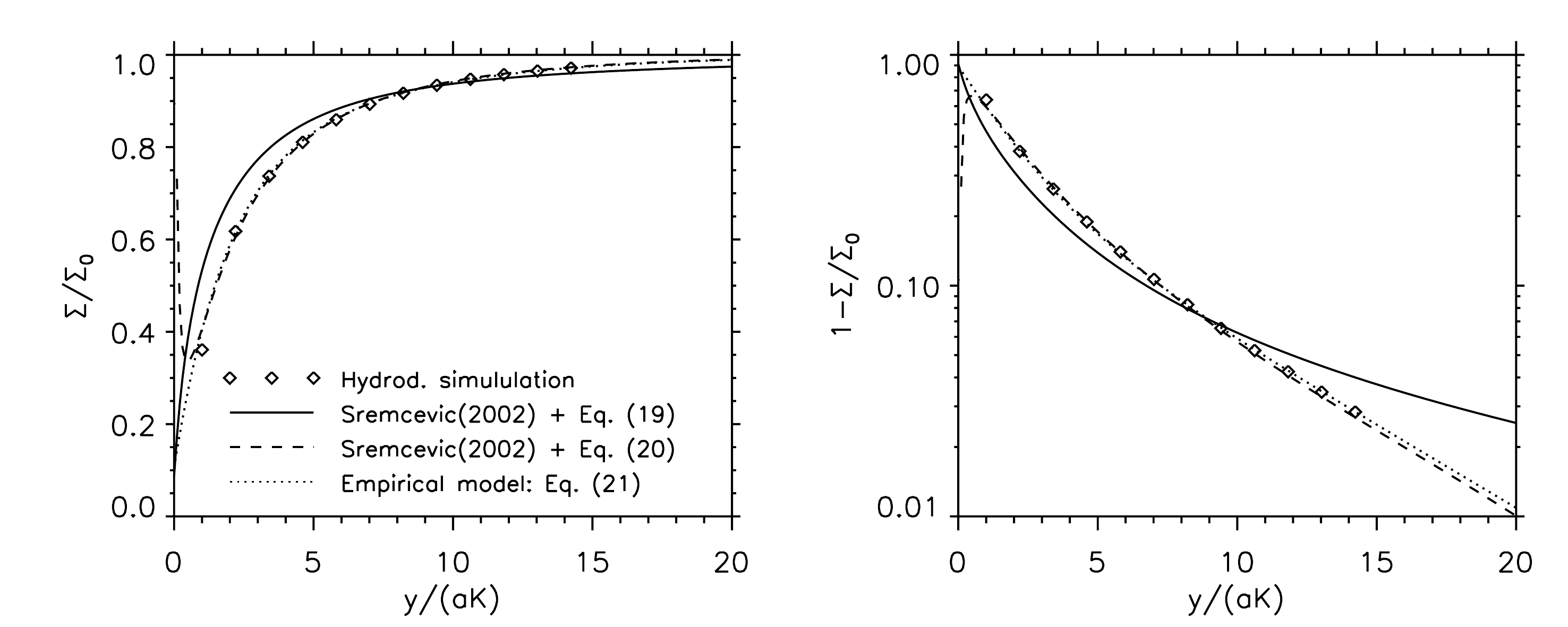}
  \caption{Comparison of the evolution of the gap minimum along the azimuth between simulations and the analytical model by \citet{SremcevicSpahnDuschl2002} - using the direct expression for the gap profile (solid line, Eq. (\ref{SGeneralSolution})) and two adjusted Dirac-$\delta$-functions (dashed line, Eq. (\ref{AnaSol})). An empirical model is plotted with a dotted line using Eq. (\ref{Empiric}).}
  \label{fig:HydroAnaModels}
\end{figure*}

Figure \ref{fig:HydroAnaModels} shows the azimuthal solution alongside the azimuthal profile from the hydrodynamic simulation, where the initial radial profile $\Sigma(x_0,y=0)$, needed to calculate the analytical solution, is taken from the hydrodynamic simulation. The analytical model does not match the simulated profile perfectly. One reason could be that \citet{SremcevicSpahnDuschl2002} assumed a fixed boundary at $x=0$. This would restrict the initial radial profile $\Sigma(x_0,y=0)$ to $x_0>0$ or $x_0<0$, but the initial density pattern exceeds this border as shown in our simulations (see Figure \ref{fig:HydroNbodyCuts}).

It has also been shown by \citet{SremcevicSpahnDuschl2002} that the source function of the density depletion can be approximated by a combination of two or three weighted Dirac $\delta$-functions, because the information of the exact initial profile gets quickly lost downstream from the moon. Thus, we additionally fit the approximation
\begin{equation}
  \frac{\Sigma(x=2,y)}{\Sigma_0} = 1 -  2.35 \cdot G(x,y,-1.29)  + 0.77 \cdot G(x,y,-4.4)
  \label{AnaSol}
\end{equation}
to the azimuthal profile. Here we keep the positions of the $\delta$-functions fixed to the mean position of the gap and the adjacent density enhancement of the general source function, but fit the prefactors to the simulation result. This approach results in a very good agreement between analytical model and simulated profile for $y > a K$. Furthermore, we find that the empirical expression with two exponential functions
\begin{equation}
	 \frac{\Sigma(x,y)}{\Sigma_0} = 1 - 0.3 \cdot \exp \left(-0.17 \frac{y}{a K} \right) 
	 			- 0.6 \cdot \exp \left(-0.55 \frac{y}{a K} \right)
	 \label{Empiric}
\end{equation}
also fits the azimuthal profile in the plotted range well.

%%%%%%%%%%%%%%%%%%%%%%%%%%%%%%%%%%%%%%%%%%%%%%%%%%%%%%%%%%%%%%%%%%%%%%%%%%
\subsection{Comparison to Cassini UVIS occultation profiles of Bl\'eriot}
%%%%%%%%%%%%%%%%%%%%%%%%%%%%%%%%%%%%%%%%%%%%%%%%%%%%%%%%%%%%%%%%%%%%%%%%%%

Since 2004 the spacecraft Cassini has been in orbit around Saturn. Among other investigations, the Ultraviolet Imaging Spectrograph (UVIS) has recorded a large number of stellar occultation observations measuring the transparency of the rings \citep{Esposito2004SSRv}. Two occultations, $\zeta$ Persei Rev 42 in April 2007 and $\alpha$ Lyrae (Vega) Rev 175 in November 2012, have been fortunate to scan across the propeller structure Bl\'eriot, located at a ring radius of 134,912 km. And indeed, the observations do show signatures consistent with a single density depletion and multiple enhancements at locations expected for Bl\'eriot's gap and wake structures \citep{Sremcevic2014DPS}. This encourages us to apply our model to Bl\'eriot and compare the results. To this aim we process UVIS occultation data in a standard manner \citep[see e.g.][]{Albers2012Icar} to obtain geometric and photometric solutions. We derive UVIS optical depth profiles at a radial resolution of 40 m smoothed with a moving average of 200 m. Then, these are compared to simulated ring surface density profiles, where we assume the simplest relation $\tau/\tau_0=\Sigma/\Sigma_0$. Due to the known, non-keplerian excess motion of the moon \citep{Tiscareno2010ApJ} the exact radial and azimuthal position of the moonlet is uncertain. The moon's position is thus a parameter in our fit, and it is adjusted so that profiles from simulation and observation coincide.

\begin{figure*}[ht!]
  \plotone{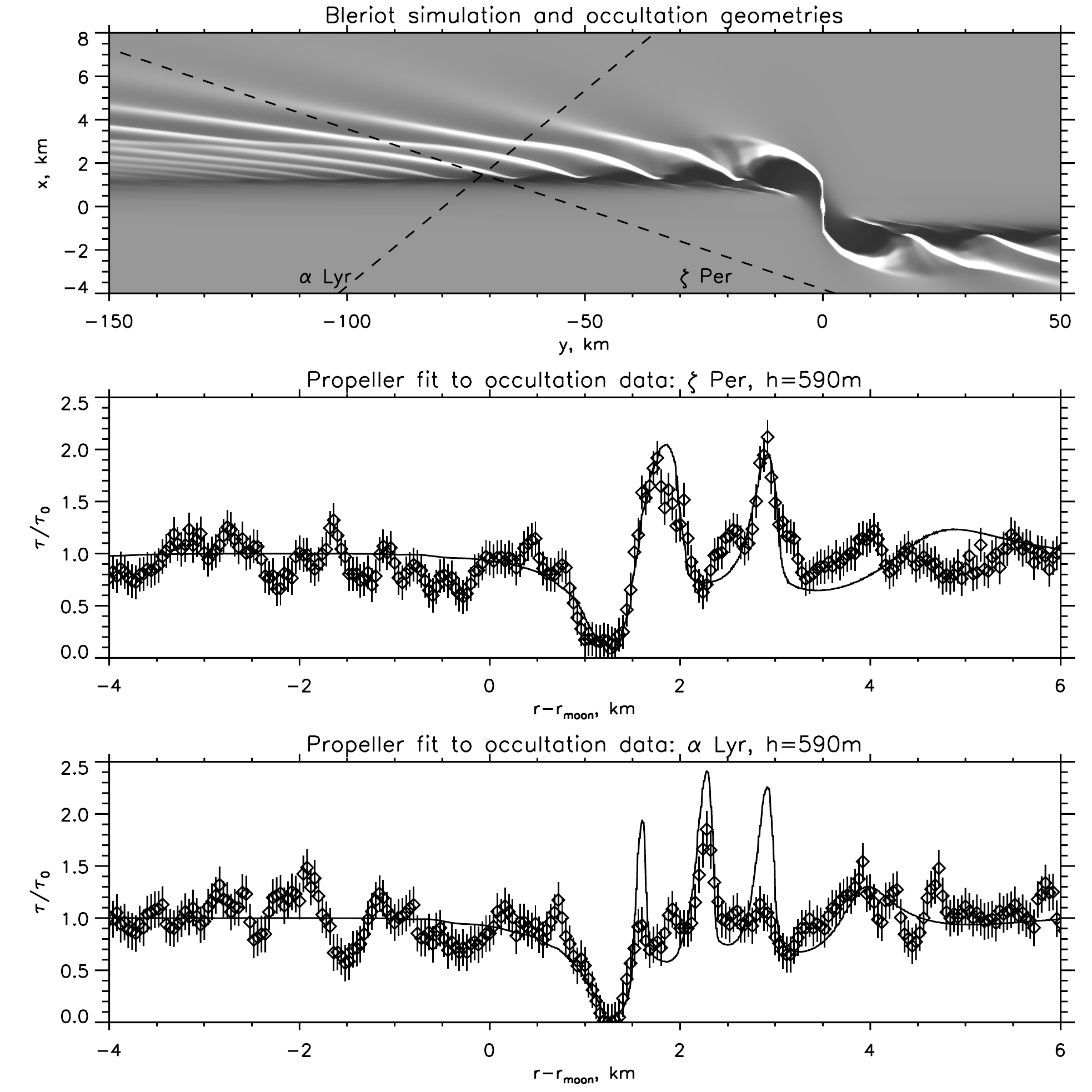}
  \caption{Comparison of a Bl\'eriot propeller simulation with two Cassini UVIS optical depth profiles. Upper panel: Simulated density and footprints of the UVIS occultation scans. Middle and lower panel: Symbols represent optical depth profiles from UVIS occultations of the stars $\zeta$ Persei and $\alpha$ Lyrae, respectively. Solid lines denote simulation results using $c_0=0.05 \, h \Omega$, $\alpha=1$, $\nu_0=0.00075 \, h^2 \Omega$, $\xi_0=0.053 \, h^2 \Omega$ and $\beta=2$.}
  \label{fig:FitUvisOccultations}
\end{figure*}

The upper panel of Figure \ref{fig:FitUvisOccultations} shows an example of a simulation result in the region of interest where the occultation paths are marked by dashed lines. In the middle and lower panel the UVIS optical depth profiles along the occultation paths are presented together with the simulation result.

Due to the uncertainty of ring parameters we perform a set of simulations and fit the data by eye. To this aim we adjust Hill radius as well as radial and azimuthal position of the moonlet in a way that gap minimum and wake maxima are at the same location for simulations and data. In order to evaluate our fit we calculate the variance of the difference between data and model
\begin{equation}
	\text{Var} = \frac{1}{N} \sum_{i=1}^{N} (\tau_{\text{d},i}-\tau_{\text{m},i})^2 
	\label{Eq:Var}
\end{equation}
where the best fitting parameters minimize the Variance. Based on the variance of the data points in an unperturbed region the uncertainty of the fit parameters is evaluated.

First, we vary the shear viscosity and fix other simulation parameters to values estimated in the Bl\'eriot region.
%expected in the Bl\'eriot region gained from theoretical estimates and measurements.
Specifically, we use Eq. (\ref{Eq:nu}) with $\beta=2$, in accordance with the expectations for the gravitational viscosity \citep{Daisaka2001}, $\alpha=1$ assuming $p=\Sigma \, c_0^2$ and $c_0 = 0.05 \, h \Omega$ \citep{Sremcevic2008DPS}. Furthermore, we use $\xi_0 = 7 \nu_0$, which is large enough to guarantee that no overstability occurs and a propeller pattern can form \citep[see e.g.][]{Schmit1995,Spahn2000Icar,Schmidt2001Icar}\footnote{There is no observational evidence for overstability in this ring region, but propellers might exist equally well in a background of overstable waves and in an unperturbed ring state.}.

\begin{figure*}[ht!]
  \plotone{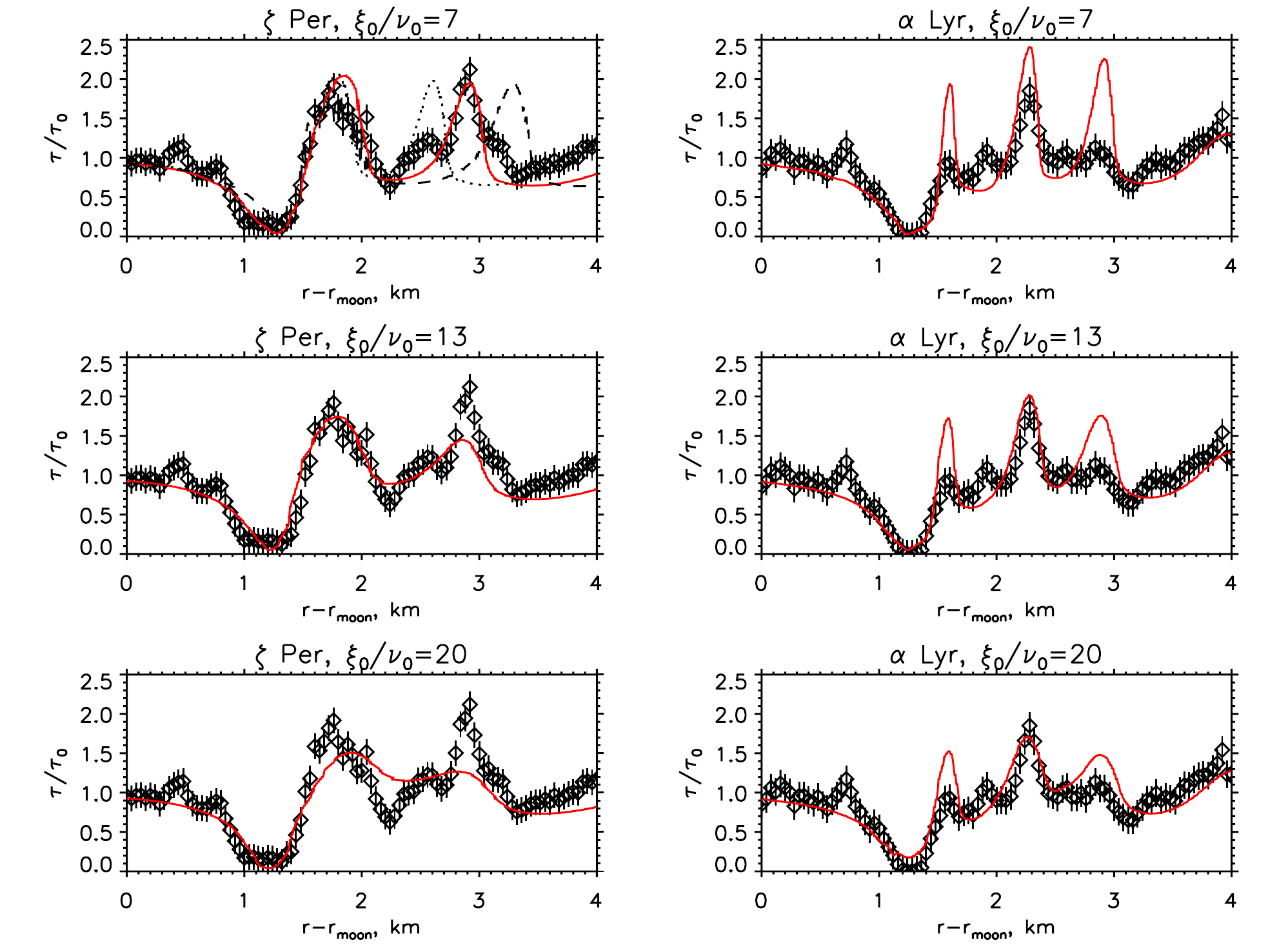}
  \caption{Comparison of two UVIS optical depth profiles to Bl\'eriot propeller simulations for different values of the bulk viscosity. The other simulation parameters are set to: $h=590 \, \text{m}$, $\nu_0=340 \, \text{cm}^2/\text{s}$ and $c_0 = 0.38  \, \text{cm} \, \text{s}^{-1}$. Dotted and dashed lines in the upper left panel denote fits with Hill radii of 490 and 690 m, respectively. Note, the small radial displacements in the data come from different fits of the moonlet positions.}
  \label{fig:ExampleFitsUvisOccultationsXi0}
\end{figure*}

The best fit of the gap region is plotted in Figure \ref{fig:ExampleFitsUvisOccultationsXi0} (first row). It is achieved for a viscosity of $\nu_0=340 \pm 120 \, \text{cm}^2 \, \text{s}^{-1}$. A Hill radius of 590 m fits the data best. The dotted and dashed line in the left panel denote profiles using Hill radii of 490 m and 690 m. This demonstrates that, although different Hill radii may fit the gap minimum and the first wake maximum, using additionally the second wake maximum for the fit constrains the Hill radius fairly tightly.

While the form of the gap is mainly determined by the shear viscosity, the form of the wake crests also depends on the action of pressure and bulk viscosity. Both fit parameters, $c_0$ and $\xi_0$, influence the wake amplitude in a similar way, so that one can not discriminate between both processes. Increasing the bulk viscosity improves the fit to the wake maxima in the $\alpha$ Lyrae profile, but underestimates the $\zeta$ Persei profile  (Figure \ref{fig:ExampleFitsUvisOccultationsXi0}, second and third row). Based on the comparison to the variance of the residual, Eq. (\ref{Eq:Var}), a reasonable range for the bulk viscosity in the hydrodynamical model is given by $7 < \xi_0/\nu_0 < 17$.
 
\begin{figure*}[ht!]
  \plotone{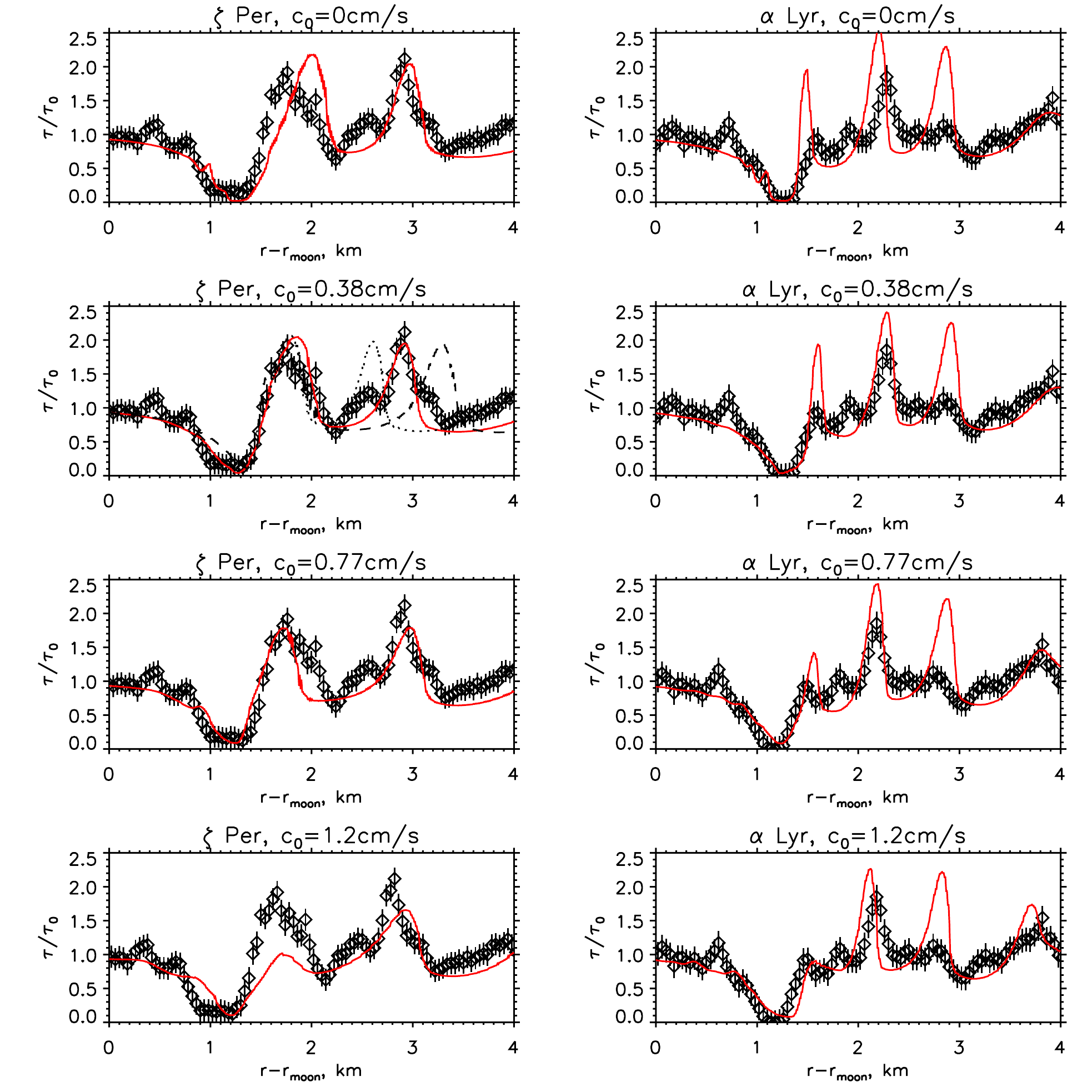}
  \caption{Comparison of two UVIS optical depth profiles to Bl\'eriot propeller simulations for different values of the dispersion velocity. The other parameters are set to: $h=590 \, \text{m}$, $\nu_0=340 \, \text{cm}^2 \, \text{s}^{-1}$ and $\xi_0 = 7 \nu_0$. Dotted and dashed lines in the second panel of the left panel column  denote fits with Hill radii of 490 and 690 m, respectively. Note, the small radial displacements in the data come from different fits of the moonlet positions.}
  \label{fig:ExampleFitsUvisOccultationsC0}
\end{figure*}

Additionally, we varied the dispersion velocity and found the best fit to the gap and wake region for $0.3 \, \text{cm} \, \text{s}^{-1} < c_0 < 1.5 \, \text{cm} \, \text{s}^{-1}$ (Figure \ref{fig:ExampleFitsUvisOccultationsC0}). In general an increased dispersion velocity leads to lower wake amplitudes.

% , but the second wake crest in the $\zeta$ Persei scan is fitted worse than when using a higher dispersion velocity

For the $\zeta$ Persei scan we find a very good agreement between observation and simulation. In the $\alpha$ Lyrae scan the gap is matched well, but only one of three predicted wake crests is observed in the data. However, the three wake crests have a radial width of about $100 \, \text{m}$ only, and therefore, might be obscured by the presence of self-gravity wakes, which have wavelengths in this size range \citep{Salo1995}. Furthermore, neglecting the finite size of the ring particles in the hydrodynamic simulations may lead to an overestimation of the wake maxima (compare to Section \ref{CompNbodyHydro}).

Measurements of the ring shear viscosity are sparse in the region of the A-ring between the Encke and Keeler gaps where Bl\'eriot orbits around Saturn. An upper limit of $\nu = 794\,\text{cm}^2 \, \text{s}^{-1}$ is given by \citet{esposito1983c}, based on the analysis of damping of the strong Janus 6:5 density wave by neglecting nonlinear effects. Recently, \citet{Tajeddine2017ApJS,Tajeddine2017Icar} derived much lower values for the shear viscosity between $10 \, \text{cm}^2 \, \text{s}^{-1}$ and $30 \, \text{cm}^2 \, \text{s}^{-1}$ in this region to explain the formation of the Keeler gap and the outer A-ring edge. However, the formation of the sharp edges are highly non-linear processes which might involve angular flux reversal \citep{Borderies1982,Borderies1989Icar} or negative diffusion \citep{Lewis2011Icar}, and thus, the results have to be taken with care. 

We consider the parameterization of \citep{Daisaka2001} for the viscosity $\nu$ of the unperturbed ring at the radial location of Bl\'eriot. This parametrization, in the form%
\footnote{Assuming a ring particle mass density of $\varrho=910\, \text{kg} \, \text{m}^{-3}$, which fits best the observations of density waves \citep{Tiscareno2007Icar}. However, other observations suggest a much lower particle density - e.g. \citet{Zhang2017Icar} and \citet{Porco2007Sci}}
\begin{equation}
\label{eq:DaisakaVis}
\nu \simeq 26 \left(\frac{r}{122\,000\,\text{km}}\right)^5\ \frac{G^2 \Sigma^2}{\Omega_0^3}\ ,
\end{equation}
was compared to values of viscosity derived from the damping of weak density waves in the inner and mid A ring \citep{Tiscareno2007Icar}, showing fairly good agreement. By using a surface mass density of $400\,\text{kg} \, \text{m}^{-2}$ \citep{Colwell2009book} and the semi-major axis of Bl\'eriot, we find a value of $\nu_0 = 160\, \text{cm}^2 \, \text{s}^{-1}$ for the viscosity of the unperturbed ring. This extrapolated value is by a factor of two smaller than the viscosity value of $\nu_0=340 \pm 120 \, \text{cm}^2 \, \text{s}^{-1}$ estimated from our simulation above, which is in turn a factor of two smaller than the upper limit by \citet{esposito1983c}. Considering that the rings are a complex system of colliding particles including inelastic collisions, fragmentation and aggregation, and that the concept of a viscosity parameter can  not account for all these processes, our viscosity value fits reasonably well to the values determined by observations of the density waves \citep{Tiscareno2007Icar}.

Measurements of the dispersion velocity in the rings are rare as well, but \citet{Sremcevic2008DPS} analyzed the dispersion relation of strong density waves in the A-ring and derived $c_0=0.3-0.5 \, \text{cm} \, \text{s}^{-1}$. In view of the large uncertainty of the method, our values for the dispersion velocity $c_0= 0.3-1.5 \, \text{cm} \, \text{s}^{-1}$ are in the same range. 

Our best fit bulk viscosity $7 < \xi_0/\nu_0 < 17$ is larger than that determined from N-body simulations using hard inelastic spheres \citep{SaloSchmidtSpahn2001}, even if one takes into account non-isothermal effects \citep{SchmidtSalo2003PhRvL}. In principle, rotational degrees of freedom of ring particles can lead to an enhancement of the bulk viscosity, if there exists a time-lag between excitations of the translational random motion and the spin temperatures \citep{ChapmanCowling1970book}. Also, if the ring particles are aggregates one might expect that a re-arrangement of the aggregate structure in collisions might affect the ratio $\xi_0/\nu_0$. In the hydrodynamic propeller model a high value of $\xi_0/\nu_0$ leads to broader, less peaked wake crests (Figure \ref{fig:ExampleFitsUvisOccultationsXi0}), and thus, to improved fits to the UVIS occultation profiles. But the observed broad wakes might physically rather result from a dense packing of ring particles at those locations where streamlines converge \citep{Borderies1985Icar}. The simple equation of state of our hydrodynamic model, Equation (\ref{Eq:PressureTensor}), does not account for this effect. Because the bulk viscosity couples to the compression of the ring material, $\nabla\cdot\vec v$, it is possible that a large value of $\xi_0$ effectively enforces solutions where $\nabla\cdot\vec v \approx 0$, i.e. the ring behaves almost incompressible. Physically, in the rings this behavior might rather be established by an equation of state $\propto (\Sigma-\Sigma_{crit})^{-1}$ that yields a diverging pressure when the granular ring matter approaches a critical density $\Sigma_{crit}$.

From Bl\'eriot's inferred Hill radius $h = (590 \pm 100) \, \text{m}$ we calculate its mass to be $M_{\text{m}} = (1.4 \pm 0.7) \cdot 10^{11} \text{kg}$. N-body simulations have shown that the propeller moons accrete material until they fill their rugby-shaped Hill sphere \citep{LewisStewart2009Icar}. This would set the density of the moon to $434 \, \text{kg}/\text{m}^3$ and its mean radius to $R_{\rm m} = 0.73 \, h = (430 \pm 70) \, \text{m}$ \citep[see also][]{Porco2007Sci}.

%%%%%%%%%%%%%%%%%%%%%%%%%%%%%%%%%%%%%%%%%%%%%%%%%%%%%%%%%%%%%%%%%%%%%%%%%
\section{Conclusions}
\label{sec:conclusion}
%%%%%%%%%%%%%%%%%%%%%%%%%%%%%%%%%%%%%%%%%%%%%%%%%%%%%%%%%%%%%%%%%%%%%%%%%

The main results of this work are:
\begin{enumerate}
\item The hydrodynamic simulations of propeller structures in Saturn's A-ring agree very well with results from N-body simulations. We neglected ring self-gravity in both approaches.
\item The hydrodynamic simulations confirm the scaling laws for the radial and azimuthal size of a propeller as predicted by \citet{SpahnSremcevic2000} and \citet{SremcevicSpahnDuschl2002}. 
\item The azimuthal evolution of the mass density downstream of the moonlet inferred from hydrodynamic simulations agrees with the analytic solution by \citet{SremcevicSpahnDuschl2002}, if the source function $\Sigma(x_0,y=0)$ is modified.
\item Comparing simulated optical depth profiles of the propeller Bl\'eriot to Cassini UVIS stellar occultation scans constrains the moonlet's Hill radius to $590 \pm 100 \, \text{m}$ with corresponding moonlet mass and diameter of ($1.4 \pm 0.7) \cdot 10^{11} \,\text{kg}$ and $(860 \pm 140)$ m, respectively.
\item The best-fit model yields a shear viscosity of the ring of $\nu = (340 \pm 120) \, \text{cm}^2 \, \text{s}^{-1}$, a ratio of the bulk viscosity to the shear viscosity in the range of $7 < \xi_0/\nu_0 < 17$, and a dispersion velocity in the range of $0.3  \, \text{cm} \, \text{s}^{-1} < c_0 < 1.5 \, \text{cm} \, \text{s}^{-1}$.
\end{enumerate}

The isothermal transport model works surprisingly well to describe the density pattern of the propeller, especially the viscous diffusion in the gap profiles. In contrast to this, \citet{Schmidt2001Icar} showed that the isothermal approximation fails to describe the overstability for the same ring conditions. Nevertheless, non-isothermal effects might influence the viscosity of the ring material, and thus, the propeller structure. Furthermore, a locally enhanced dispersion velocity would also cause an enhanced local thickness of the ring, as it has been observed for the propeller Earhart around Saturn's vernal equinox \citep{Hoffmann2013ApJ,Hoffmann2015Icar}. However, UVIS occultations of the stars $\zeta$ Persei and $\alpha$ Lyrae were observed at a moderate elevation angle of 38 and 35 degrees, respectively, and thus, the vertical ring structure should be of minor importance in these two cases. Moreover, including ring self-gravity in the model would allow to investigate the influence of self-gravity wakes on the propeller, especially in the moonlet wake region; but this is left for future work.

\section*{Acknowledgement}
We thank the reviewer Matthew S. Tiscareno for his detailed report and the helpful comments. This work has been supported by the Deutsches Zentrum f{\"u}r Luft-und Raumfahrt (OH 1401), the Deutsche Forschungsgemeinschaft (Sp 384/28-1, Ho5720/1-1), and the Cassini project.

%\input{/Users/Martin/ARBEIT/SOURCE/BIBLIOGRAPHY/journals2}
%\input{bibliography}
%\bibliographystyle{/Users/Martin/ARBEIT/SOURCE/BIBLIOGRAPHY/pss}
%\bibliography{/Users/Martin/ARBEIT/SOURCE/BIBLIOGRAPHY/martins_ringe}

\end{document}